\documentclass[aps,prl,amsmath,amssymb,superscriptaddress,preprintnumbers,twocolumn,10pt,nofootinbib]{revtex4-1}
\usepackage{CJKutf8}
\usepackage{graphicx}
\usepackage{dcolumn}
\usepackage{bm}
\usepackage{amssymb}
\usepackage{latexsym}
\usepackage{booktabs}
\usepackage{amsmath}
\usepackage{multirow}
\usepackage{url}
\usepackage{footnote}
\usepackage{float}
\usepackage{threeparttable}
\usepackage{subfigure}
\usepackage[bottom]{footmisc}

\usepackage[normalem]{ulem}
\usepackage{color}
\usepackage{array}
\usepackage{enumerate}
\usepackage{adjustbox}
\usepackage{makecell}
\usepackage{diagbox}
\usepackage{epstopdf}
\usepackage{epsfig}
\usepackage{longtable}
\usepackage{supertabular}
\usepackage{algorithm}
\usepackage{pifont}
\usepackage{algorithmic}
\usepackage{changepage}
\usepackage{setspace}
\usepackage{lineno}
\usepackage[colorlinks=true, linkcolor=blue, citecolor=blue]{hyperref}

\begin{document}

\title{Geometric unification of timelike orbital chaos and phase transitions in black holes}

\author{Shi-Hao Zhang}
\affiliation{Liaoning Key Laboratory of Cosmology and Astrophysics, College of Sciences, Northeastern University, Shenyang 110819, China}

\author{Zi-Yuan Li}
\affiliation{School of Physics, Nankai University, Tianjin 300071, China}

\author{Jing-Fei Zhang}
\affiliation{Liaoning Key Laboratory of Cosmology and Astrophysics, College of Sciences, Northeastern University, Shenyang 110819, China}

\author{Xin Zhang}\thanks{Corresponding author}\email{zhangxin@mail.neu.edu.cn}
\affiliation{Liaoning Key Laboratory of Cosmology and Astrophysics, College of Sciences, Northeastern University, Shenyang 110819, China}
\affiliation{MOE Key Laboratory of Data Analytics and Optimization for Smart Industry, Northeastern University, Shenyang 110819, China}
\affiliation{National Frontiers Science Center for Industrial Intelligence and Systems Optimization, Northeastern University, Shenyang 110819, China}

\begin{abstract}

The deep connection between black hole thermodynamics and spacetime geometry remains a central focus of general relativity. While recent studies have revealed a precise correspondence for null orbits, given by $K = -\lambda^2$ between the Gaussian curvature $K$ and the Lyapunov exponent $\lambda$, its validity for timelike orbits had remained unknown. Our work introduces the massive particle surface (MPS) framework and constructs a new geometric quantity $\mathcal{G}$. We demonstrate that $\mathcal{G} \propto -\lambda^2$ on unstable timelike orbits, thus establishing the geometry-dynamics correspondence for massive particles. Crucially, near the first-order phase transition of a black hole, $\mathcal{G}$ displays synchronized multivalued behavior with the Lyapunov exponent $\lambda$ and yields a critical exponent $\delta=1.0244$. Our results demonstrate that spacetime geometry encodes thermodynamic information, opening a new pathway for studying black hole phase transitions from a geometric perspective.

\end{abstract}

\maketitle

\section{Introduction}
Black holes, remarkable consequences of general relativity, exhibit rich thermodynamic phenomena \cite{Hawking:1975,Bekenstein:1973}, including first-order phase transitions analogous to van der Waals fluids \cite{Kubiznak:2012}. This finding challenges the conventional boundaries between gravity, thermodynamics, and statistical physics, raising a central question: How are the thermodynamic properties of black holes fundamentally encoded in the geometric structure of spacetime?

Recent studies have offered a promising perspective on this question. It has been discovered that the Lyapunov exponent $\lambda$, which characterizes the chaotic motion of massless particles on unstable null orbits around black holes, satisfies an exact correspondence with the Gaussian curvature $K$ of the optical metric \cite{Gallo:2025}: $K = -\lambda^2$. Furthermore, during the first-order phase transition of black holes, both quantities exhibit characteristic multivalued behavior \cite{Guo:2022,Yang:2023,Lyu:2024,Kumara:2024,Du:2025,Shukla:2024,Gogoi:2024,Chen:2025,Karthik:2025,Awal:2025,Yang:2025,Zhang:2025}. These findings suggest that the key information linking dynamics and thermodynamics is inherently encoded in spacetime geometry.

However, this established geometric-dynamic correspondence encounters a fundamental challenge when generalized to massive particles. Since the motion of massive particles is described by a Jacobi metric that depends on both energy and mass, the Gaussian curvature of their orbits no longer maintains its exact relation with the Lyapunov exponent. This leaves a critical gap: Is the chaotic behavior of massive particles also encoded in spacetime geometry? Consequently, does spacetime geometry remain a valid probe for first-order phase transitions in the timelike case?

This work answers both questions by introducing the massive particle surface (MPS) framework. We construct a new geometric quantity $\mathcal{G}$ to demonstrate that, for neutral massive particles on unstable timelike orbits, the relation $\mathcal{G} \propto -\lambda^2$ holds. This result successfully extends the geometry-dynamics correspondence from massless to massive particles. Crucially, in the spinodal region of the black hole first-order phase transition, both $\mathcal{G}$ and the Lyapunov exponent $\lambda$ simultaneously exhibit multivalued behavior. By examining the critical behavior of these quantities near the phase transition point, we find that their critical exponents deviate from the value $1/2$ observed in the Reissner–Nordström black hole. This indicates that regular black holes exhibit richer critical behavior in their geometry and dynamics than their singular counterparts. Our work establishes spacetime geometry as a unified foundation connecting dynamic chaos and thermodynamic phase transitions in black holes, thereby enabling a geometric perspective for probing their thermodynamics. We set $G=c=k_B=\hbar=1$ in this paper.

\section{Timelike Unstable Orbits and Lyapunov Exponent}
The Lyapunov exponent $\lambda$ quantifies the chaotic behavior of particle motion on timelike circular orbits near a black hole. 
First, we briefly review the derivation of this exponent for unstable timelike orbits.

Consider a 4-dimensional static spherically symmetric metric
\begin{align}
ds^{2}=-fdt^{2}+\frac{1}{g}dr^{2}+r^{2}d\Omega^{2}\label{metric 01},
\end{align}
where $d\Omega^{2}=r^2 \left( d\theta^2 + \sin^2 \theta d\phi^2 \right)$ is the unit 2-sphere, and $f,~g$ are $C^P$ smooth functions of $r$ with $P\geq2$. For timelike orbits, the Lagrangian is
\begin{align}
\mathcal{L} = g_{\mu\nu} \dot{x}^\mu \dot{x}^\nu = -1,\label{Lagrangian}
\end{align}
where the dot denotes a derivative with respect to affine parameter. The two conserved quantities of the particle, its energy $\mathcal{E}$ and angular momentum $L$, are given by
\begin{align}
\quad -\mathcal{E} =  \frac{\partial \mathcal{L}}{\partial \dot{t}} = -f \dot{t}, ~~L =  \frac{\partial \mathcal{L}}{\partial \dot{\phi}} = r^2 \dot{\phi}.\label{LE}
\end{align}
We expand the Lagrangian around the unstable circular orbit located at $r=r_0$, and restrict our attention to the metric in the equatorial plane ($\theta= \frac{\pi}{2}$), which leads to
\begin{align}
(\dot{\varepsilon})^2 - \frac{1}{2} V''_{\text{eff}} \varepsilon^2 = 0,
\end{align}
the prime notation denotes differentiation with respect to the radial coordinate $r$, and define the Lyapunov exponent as
\begin{align}
\lambda = \sqrt{\frac{1}{2(\dot{t})^2} V_\text{eff}''}\bigg|_{r=r_0}.
\end{align}
With
\begin{align}
V_{\text{eff}}= g \left( \frac{E^2}{f} - \frac{L^2}{r^2}-1\right),
\end{align}
where $r_0$ is determined by the conditions $V_\text{eff}(r_0) = V_\text{eff}'(r_0) = 0$, and $\varepsilon=r-r_{0}$. When $f=g$, we obtain
\begin{align}
\lambda^2 = \frac{1}{2} \left[ -3 \frac{f f'}{r_0} + 2 (f')^2 - f f'' \right] \bigg|_{r=r_0}.\label{lambda}
\end{align}

In the subsequent analysis, we seek to establish a connection between a geometric quantity and the Lyapunov exponent of Eq.~(\ref{lambda}).

\section{Jacobi Metric and the Limitation of Gaussian Curvature}
For a $4$-dimensional spherically symmetric metric as in Eq.~(\ref{metric 01}), a precise correspondence between orbital dynamics and geometry for unstable null orbits \cite{Gallo:2025}
\begin{align}
K(r_{LR}) = -\lambda^2(r_{LR}).\label{lambdaK}
\end{align}
Here, $r_{LR}$ denotes the radius of the unstable null circular orbit (light ring). We need to investigate whether the Gaussian curvature is related to the Lyapunov exponent on unstable timelike circular orbits.

As a generalization of the optical metric, the Jacobi metric provides an effective description of geodesics for massive particles moving freely in spacetime \cite{Gibbons:2015qja,Arganaraz:2021fwu,Bermudez:2024bfi}. For a neutral, free particle of mass $m$ and energy $\mathcal{E}$ in a spherically symmetric spacetime, the Jacobi metric $J_{ij}$ is given by
\begin{align}
J_{ij} dx^i dx^j = \frac{\mathcal{E}^2 - m^2 f}{f} \left( \frac{1}{g} dr^2 + r^2 d\Omega^2 \right).\label{JM}
\end{align}
This metric, which satisfies the normalization condition $J^{ij} J_{ij} = 1$, describes the geodesic motion of a particle with mass $m$ on a surface of constant energy $\mathcal{E}$. Note that for the null case $m=0$, the Jacobi metric reduces to the optical metric. In this limit, Eq.~(\ref{JM}) can be cast into
\begin{align}
J_{ij} dx^i dx^j = \mathcal{E}^2 g_{ij}^{OP} dx^i dx^j.
\end{align}
The optical metric $g_{ij}^{OP}$ is defined as
\begin{align}
g_{ij}^{OP} dx^i dx^j = \frac{1}{f} \left( \frac{1}{g} dr^2 + r^2 d\Omega^2 \right),
\end{align}
which coincides exactly with the optical metric obtained by setting $ds^2=0$ in the spacetime metric Eq.~(\ref{metric 01}).

The Gaussian curvature for the unstable timelike circular orbits $r_0$ in the metric given by Eq.~(\ref{JM}) is obtained as follows. For a 2-dimensional Riemannian manifold in orthogonal coordinates, its Gaussian curvature is given by
\begin{align}
K = -\frac{1}{2} \frac{1}{\sqrt{g_{rr} g_{\phi\phi}}} \frac{d}{dr} \left( \frac{g'_{\phi\phi}}{\sqrt{g_{rr} g_{\phi\phi}}} \right).
\end{align}
On the circular orbit $r_0$, where the geodesic curvature vanishes, ${\kappa}_g = \left[ \frac{1}{2\sqrt{g_{rr}}} \frac{\partial \ln(g_{\phi\phi})}{\partial r} \right] \bigg|_{r_0} = 0$, this expression simplifies to \cite{Bermudez:2024bfi}
\begin{widetext}
\begin{align}
K_{\text{cric}} = \left\{-\frac{1}{2r\mathcal{E}^2\left(1-\frac{m^2}{\mathcal{E}^2}f\right)^2}\left[rf'^2\left(\frac{1-\frac{2m^2}{\mathcal{E}^2}f}{1-\frac{m^2}{\mathcal{E}^2}f}\right)-f(f'+rf'')\right]\right\}\bigg|_{r_0}.\label{K}
\end{align}
\end{widetext}

A comparison of Eq.~(\ref{K}) with Eq.~(\ref{lambda}) reveals that for unstable timelike circular orbits, the Gaussian curvature no longer admits a straightforward relation with the Lyapunov exponent.

This result highlights a fundamental distinction between massless and massive particles. For massless particles, the optical metric and thus the Gaussian curvature of null orbits are independent of particle energy and mass. However, when considering unstable orbits formed by massive particles, the Jacobi metric depends explicitly on particle energy $\mathcal{E}$ and mass $m$. Consequently, the Gaussian curvature also becomes dependent on $\mathcal{E}$ and $m$, which explains why a correspondence of the form given in Eq.~(\ref{lambdaK}) fails in the timelike case.

Nevertheless, as shown in the following analysis, although the Gaussian curvature no longer maintains a quantitative relation with the Lyapunov exponent in the timelike case, another geometric quantity does exhibit a definitive correspondence with it.

\section{MPS Geometry}
We briefly review the fundamental concepts and properties of the MPS defined in Refs.~\cite{Kobialko:2022uzj,Bogush:2023ojz,Bogush:2024fqj}, and establish the connection between the associated geometric quantity $\mathcal{G}$ and the Lyapunov exponent.

The MPS is defined as follows. Consider a Lorentzian manifold $\mathcal{M}$ of dimension $d \geq 4$ with a Killing vector $k^\alpha$. A massive particle surface is an immersed, timelike hypersurface $\Sigma$ of $\mathcal{M}$ such that, for every point $p \in \Sigma$ and every tangent vector $v^\alpha \in T_p \Sigma$ satisfying $v^\alpha k_\alpha = -\mathcal{E}$ and $v^\alpha v_\alpha = -m^2$, there exists a worldline $x^\mu$ of a particle with mass $m$ and total energy $\mathcal{E}$ for which $v^\mu(0) = v^\mu\big|_{p}$ and $x^\mu \subset \Sigma$ \cite{Kobialko:2022uzj}. Here, the Killing vector $k^\alpha$ satisfies the Killing equation
\begin{align}
\nabla_{(\mu}k_{\nu)}=0.
\end{align}

Consider a neutral test particle of mass $m$ whose worldline in $\mathcal{M}$ is governed by
\begin{align}
v^\mu \nabla_\mu v^\nu = 0,~\quad v^\mu v_\mu = -m^2.
\end{align}
The particle energy is defined as $\mathcal{E} = -k_\mu v^\mu$, where $v^\mu = \frac{dx^\mu}{ds}$ is the four-velocity, $s$ is an affine parameter, and $v^\mu$ is tangent to the MPS. Consequently, energy is conserved along the worldline
\begin{align}
\frac{d\mathcal{E}}{ds} = 0.
\end{align}
At each point $p$ on the massive particle surface $\Sigma$, there exists a family of fixed four-velocities $v^\mu \in T_p\Sigma$ that satisfy both the energy constraint $v^\mu k_\mu = -\mathcal{E}$ and the normalization condition $v^\mu v_\mu = -m^2$. Each such four-velocity defines a specific worldline for a particle with the fixed energy $\mathcal{E}$ and mass $m$. The union of all these worldlines through all points of $\Sigma$ constitutes the $(d-1)$-dimensional timelike hypersurface $\Sigma$ itself.

To characterize the geometry of $\Sigma$, we introduce a unit normal vector $n^\mu$. The first fundamental form (induced metric) of $\Sigma$ is given by
\begin{align}
h_{\mu\nu} = g_{\mu\nu} - n_\mu n_\nu,
\end{align}
or equivalently expressed as a projection tensor
\begin{align}
h^\mu_\nu = \delta^\mu_\nu - n^\mu n_\nu.
\end{align}
The extrinsic curvature (second fundamental form) of $\Sigma$ is then given by
\begin{align}
\chi_{\mu\nu} = h^\alpha_\mu h^\beta_\nu \nabla_\alpha n_\beta.
\end{align}
The Killing vector can be projected onto the hypersurface $\Sigma$ and decomposed as
\begin{align}
k^\mu = \tilde{k}^\mu + k_+ n^\mu,~\quad \tilde{k}^\mu n_\mu = 0,\label{decomposition}
\end{align}
where $\tilde{k}^\mu$ is the tangential component of ${k}^\mu$ on $\Sigma$, and $k_+ n^\mu$ is the normal component. The four-velocity admits a similar decomposition
\begin{align}
v^\mu = A \tilde{k}^\mu + B u^\mu,\label{decomposition2}
\end{align}
where $u^\mu$ is a tangent vector on $\Sigma$ orthogonal to both $\tilde{k}^\mu$ and $n_\mu$, i.e., $\tilde{k}_\mu u^\mu = n_\mu u^\mu = 0$, and $A$, $B$ are coefficients determined below.
By considering the normalization condition $v^\mu v_\mu = -m^2$ and $v^\mu k_\mu = -\mathcal{E}$, we obtain
\begin{align}
A = -\frac{\mathcal{E}}{\tilde{k}^2},~B = 1.
\end{align}
Substituting these into Eq.~(\ref{decomposition2}) yields
\begin{align}
u^2 = -m^2 - \frac{\mathcal{E}^2}{\tilde{k}^2}.
\end{align}
Furthermore, since $\tilde{k}^\mu$ is timelike ($\tilde{k}^2<0$), it follows that
\begin{align}
0 < \left| \tilde{k}^\mu \tilde{k}_\mu \right| \leq \frac{\mathcal{E}^2}{m^2}.\label{uneq}
\end{align}
In the massless case ($m=0$), the right-hand side of Eq.~(\ref{uneq}) diverges, which reflects the conformal invariance of null geodesic equations. For $m>0$, Eq.~(\ref{uneq}) describes a constraint on the motion of massive particles over the MPS.

It has been shown in Ref.~\cite{Kobialko:2022uzj} that for such a hypersurface, the extrinsic curvature can always be expressed as
\begin{align}
\chi_{\alpha\beta} = \frac{\chi_\tau}{d-2} H_{\alpha\beta},\label{chi}
\end{align}
where
\begin{align}
\chi_\tau = \frac{d-2}{H} \chi^\alpha_\alpha,~\quad H = H^\alpha_\alpha,~\quad H_{\alpha\beta} = h_{\alpha\beta} + \frac{m^2}{\mathcal{E}^2} \tilde{k}_\alpha \tilde{k}_\beta.
\end{align}

This is the special uncharged case of Ref.~\cite{Kobialko:2022uzj}. On the MPS, consider an orthonormal basis constructed from the timelike Killing vector $\tilde{k}^\mu$ on $\Sigma$ and a set of $d-2$ linearly independent vectors $\tau^\alpha_{(i)}$ defined on $\Sigma$, satisfying
\begin{align}
\tau^\alpha_{(i)} \tilde{k}_\alpha = 0.
\end{align}
It can be shown that this construction leads to a central relation known as the master equation \cite{Kobialko:2022uzj}
\begin{align}
\mathcal{E} = \pm m \sqrt{\frac{\tilde{k}^2 \chi_\tau}{W}},~W = -\chi_\tau - (d - 2) \tilde{k}^{-2} \tilde{k}^\alpha n^\beta \nabla_\alpha \tilde{k}_\beta.\label{ME}
\end{align}
The master equation reflects the fact that the hypersurface $\Sigma$ is partially umbilical. Consequently, neutral massive particles on $\Sigma$ possess specific energy $\mathcal{E}$ and mass $m$ satisfying this equation. Since Eq.~(\ref{ME}) combines $\mathcal{E}$ and $m$ with spacetime curvature, it enables the derivation of several important relations.

Now consider a 4-dimensional static spherically symmetric metric as in Eq.~(\ref{metric 01}). For this metric, we obtain
\begin{align}
\chi_\tau = \sqrt{g} \partial_r \ln h,~W = - \sqrt{g} \partial_r \ln \left( \frac{h}{f} \right).
\end{align}
Substituting these into Eq.~(\ref{ME}) leads to
\begin{align}
\frac{\mathcal{E}^2}{m^2} = \frac{f^2 h'}{h' f - h f'}\label{ME02}.
\end{align}
For massless particles
\begin{align}
h'f - h f' = 0.\label{LR}
\end{align}
Solving Eq.~(\ref{LR}) under this condition yields the radius $r_{LR}$ of the null unstable circular orbit, which is fully consistent with the result obtained via the dynamical approach. This confirms that in the massless limit, the MPS naturally reduces to the photon surface. For $h=r^2$, one can find
\begin{widetext}
\begin{align}
\frac{d}{dr} \left( \frac{\tilde{k}^2 \chi_\tau}{W} \right) = \frac{d}{dr} \left( \frac{\mathcal{E}^2}{m^2} \right) = \frac{2fr}{(2f - rf')^2} \left( 3\frac{ff'}{r} - 2f'^2 + ff'' \right) .\label{ME03}
\end{align}
\end{widetext}
Comparing Eq.~(\ref{ME03}) with Eq.~(\ref{lambda}) leads to
\begin{align}
\mathcal{G} = -\frac{4fr}{(2f - rf')^2} \lambda^2,\label{G}
\end{align}
where the geometric quantity $\mathcal{G}$ is defined as
\begin{align}
\mathcal{G} = \frac{d}{dr} \left( \frac{\tilde{k}^2 \chi_\tau}{W} \right).
\end{align}
As shown in Eq.~(\ref{G}), we establish that the geometric quantity $\mathcal{G}$, related to the extrinsic curvature of the MPS, is proportional to the square of the Lyapunov exponent. This successfully generalizes the correspondence between geometry and dynamics from the null case in Eq.~(\ref{lambdaK}) to the timelike case. This demonstrates that spacetime geometry encodes the chaotic behavior of the black hole–particle system. Moreover, as will be shown in the following result, it also contains thermodynamic information.

It is worth discussing that substituting Eq.~(\ref{ME02}) into Eq.~(\ref{K}) for timelike orbits yields the relation $K_{\text{cric}} = -\left[\frac{4f^2}{r^2\mathcal{E}^2(f')^2}\right]\bigg|_{r_0}\lambda^2$. This might suggest a mathematical proportionality $K_{\text{cric}} \propto \lambda^2$, although it has a coefficient dependent on $\mathcal{E}$. However, this operation constitutes a conceptual mismatch. Although the unstable timelike orbit at $r_0$ is a subset of the MPS, the Jacobi metric is an effective geometry constructed from particle geodesics, and its curvature is a geometric quantity of this effective space. In contrast, the MPS is a hypersurface immersed in the physical spacetime, and its master equation describes an inherent geometric constraint. Substituting a relation from this spacetime geometry into the calculation of an effective geometric quantity conflates two distinct geometric frameworks.

These considerations clarify the fundamental limitation that for the timelike case, any relation between the Gaussian curvature and the Lyapunov exponent is inherently dependent on $\mathcal{E}$, and thus cannot offer a universal correspondence. Instead, our work establishes that the geometric quantity $\mathcal{G}$, defined within the framework of spacetime geometry, provides a direct and universal correspondence $\mathcal{G} \propto -\lambda^2$. This successfully generalizes the geometry-dynamics correspondence from the null to the timelike case.

\section{Results}
Eq.~(\ref{G}) is now employed to demonstrate the multivalued behavior of $\mathcal{G}$ near black hole first-order phase transition points, thereby showing its effectiveness in probing such phase transitions in the timelike case. We consider the Hayward-Letelier-AdS black hole as a specific example.

We begin by reviewing relations in black hole thermodynamics. For a $4$-dimensional spherically symmetric black hole described by the metric in Eq.~(\ref{metric 01}), when $f=g$ and $h=r^2$, the Hawking temperature is given by
\begin{align}
T = \frac{f'}{4\pi} \bigg|_{r_+}.
\end{align}
Here, $r_+$ denotes the event horizon radius defined by $f(r_+)=0$. The Gibbs free energy is given by
\begin{align} 
F = M-TS,
\end{align}
where $M$ is the ADM mass of the black hole and $S =\pi r_+^2$ is the entropy.  When a black hole undergoes a first-order phase transition, the $F(T)$ curve exhibits a swallowtail structure, corresponding to the existence of three black hole solutions: large, intermediate, and small black holes. The critical point is an inflection point determined by the conditions
\begin{align}
\frac{\partial T}{\partial r_+} = \frac{\partial^2 T}{\partial r_+^2} = 0.
\end{align}
If these critical conditions are not met, only a single black hole solution exists.

The metric function of the Hayward-Letelier-AdS black hole is given by \cite{Kumar:2024qon}
\begin{align}
f_{HL}(r) = 1 - \frac{2Mr^2}{g^3 + r^3} + \frac{r^2}{\ell^2}-a,\label{fhl}
\end{align}
where $g$ is the magnetic monopole charge, $\ell$ is the AdS radius, and the string-cloud parameter $a$ arises from the string-cloud term in action \cite{Kumar:2024qon}. When $a=0$, the metric reduces to the Hayward–AdS black hole. Since $a$ is a dimensionless constant, we introduce the following scaling
\begin{widetext}
\begin{align}
\tilde{r}_+ = \frac{r_+}{\ell}, \quad \tilde{g} = \frac{g}{\ell}, \quad \tilde{M} = \frac{M}{\ell}, \quad \tilde{F}_{HL} = \frac{F_{HL}}{\ell}, \quad \tilde{T}_{HL} = T_{HL}\ell.
\end{align}
\end{widetext}

\begin{figure*}
	\begin{minipage}{1\hsize}
		\begin{center}
			
            \subfigure[$\tilde{F}_{HL}-\tilde{T}_{HL}$]{
				\label{HLFT}
				\includegraphics*[scale=0.23]{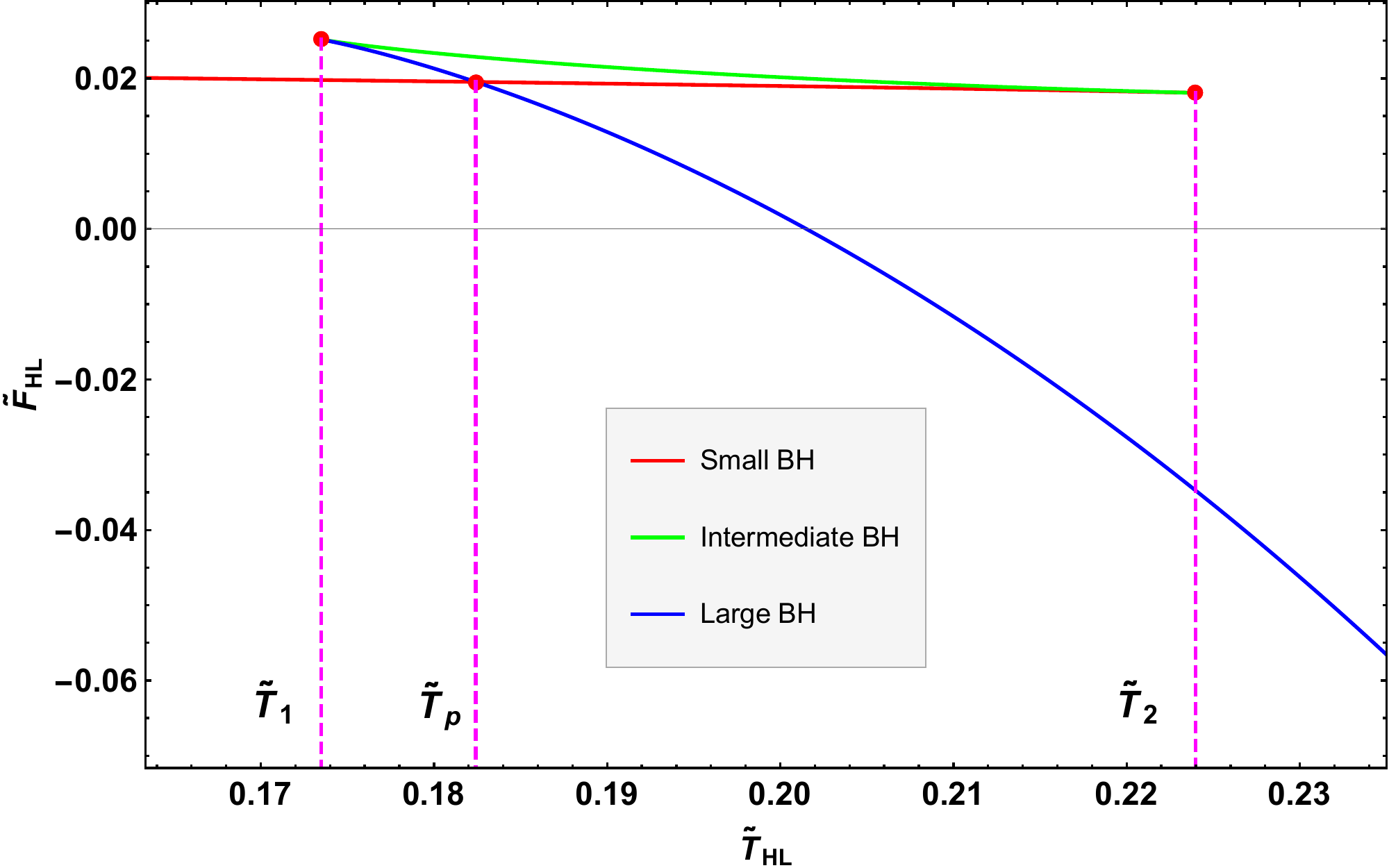}}
            \subfigure[$\tilde{F}_{HL}-\tilde{T}_{HL}$]{
				\label{HLKT}
				\includegraphics*[scale=0.25]{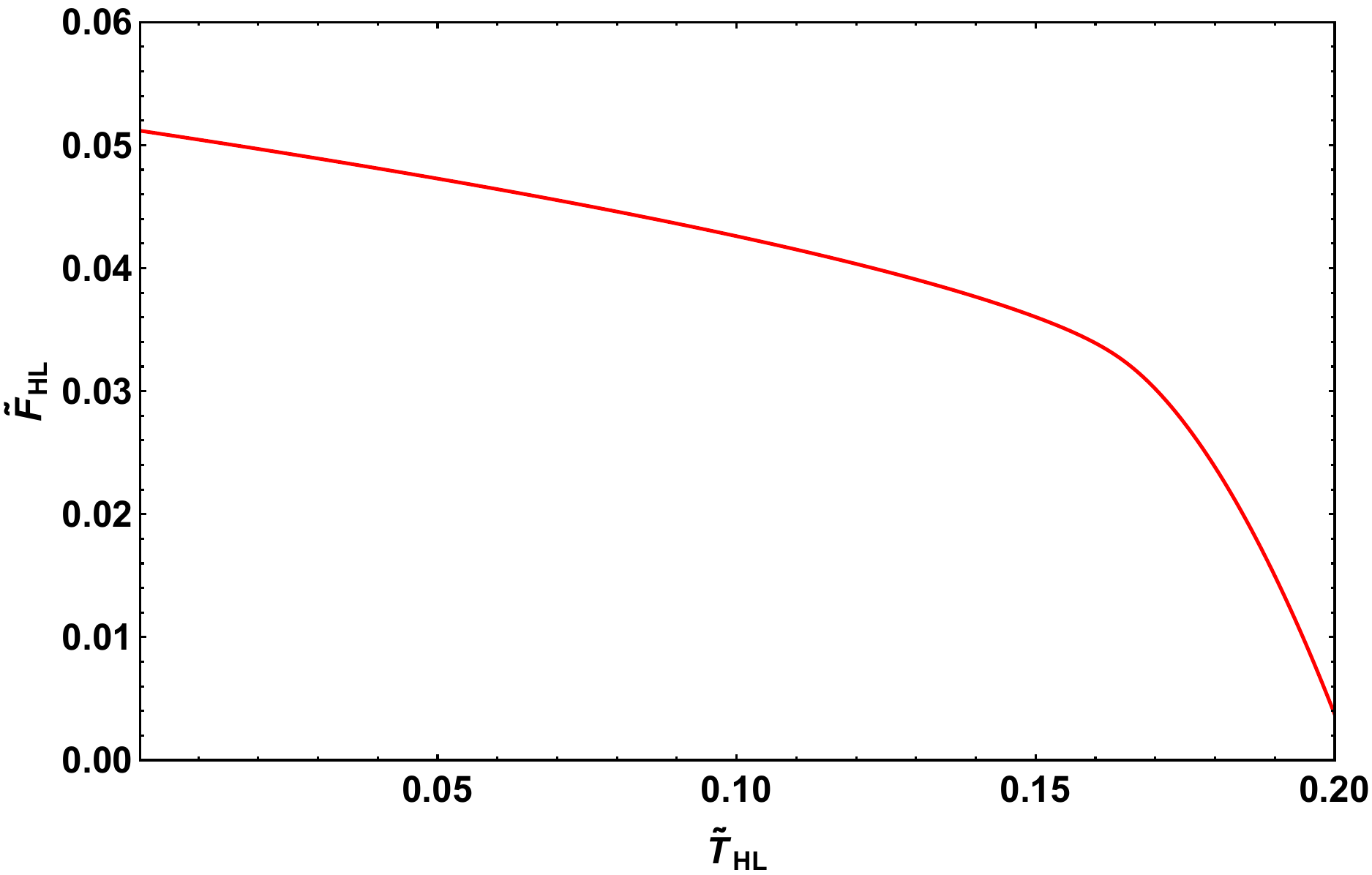}}    
			\subfigure[$\lambda_{HL}-\tilde{T}_{HL}$ (Null)]{
				\label{HLNullLT}
				\includegraphics*[scale=0.23]{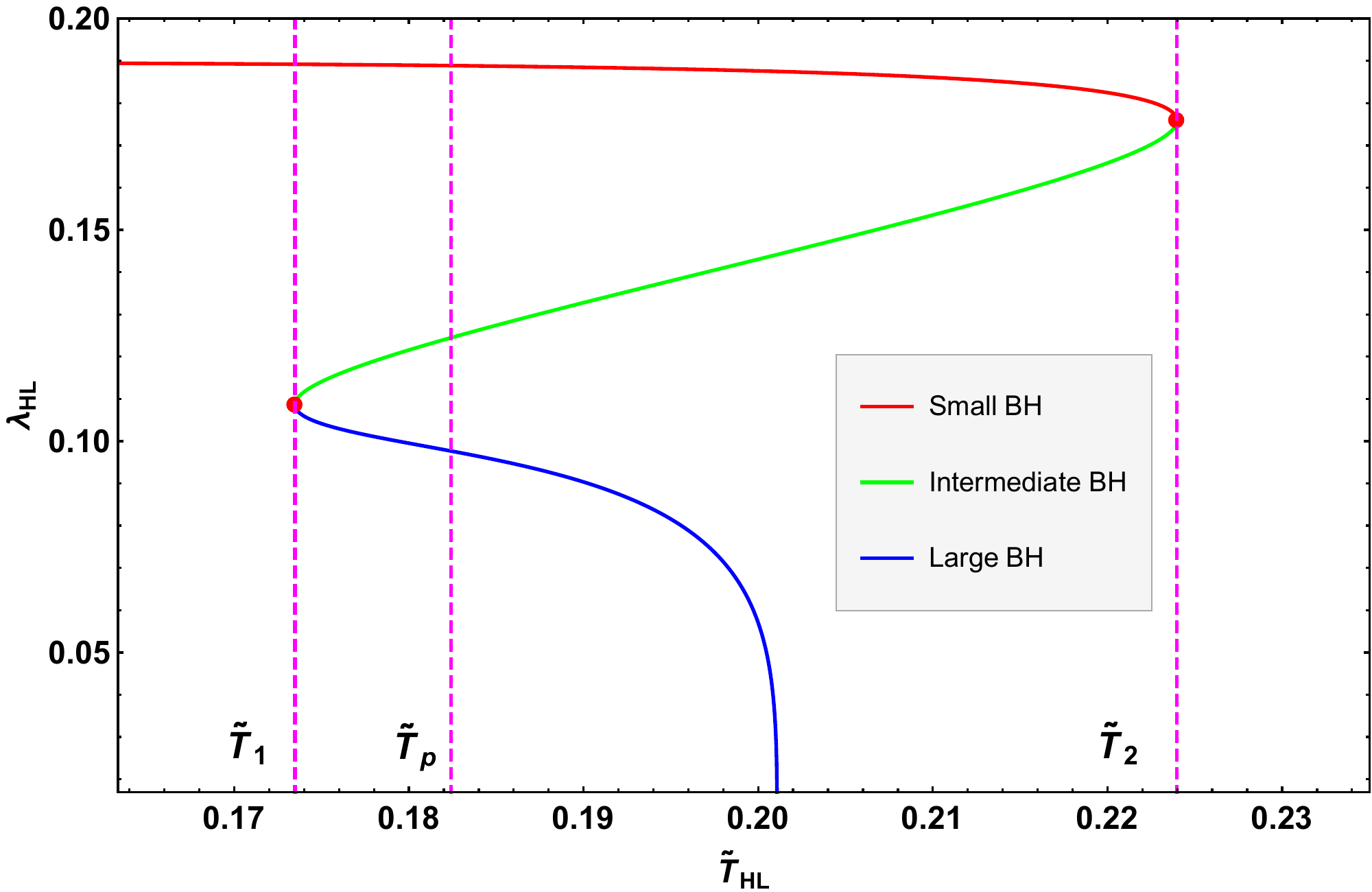}}
			\subfigure[$\lambda_{HL}-\tilde{T}_{HL}$]{
				\label{HLTimelikeLT}
				\includegraphics*[scale=0.25]{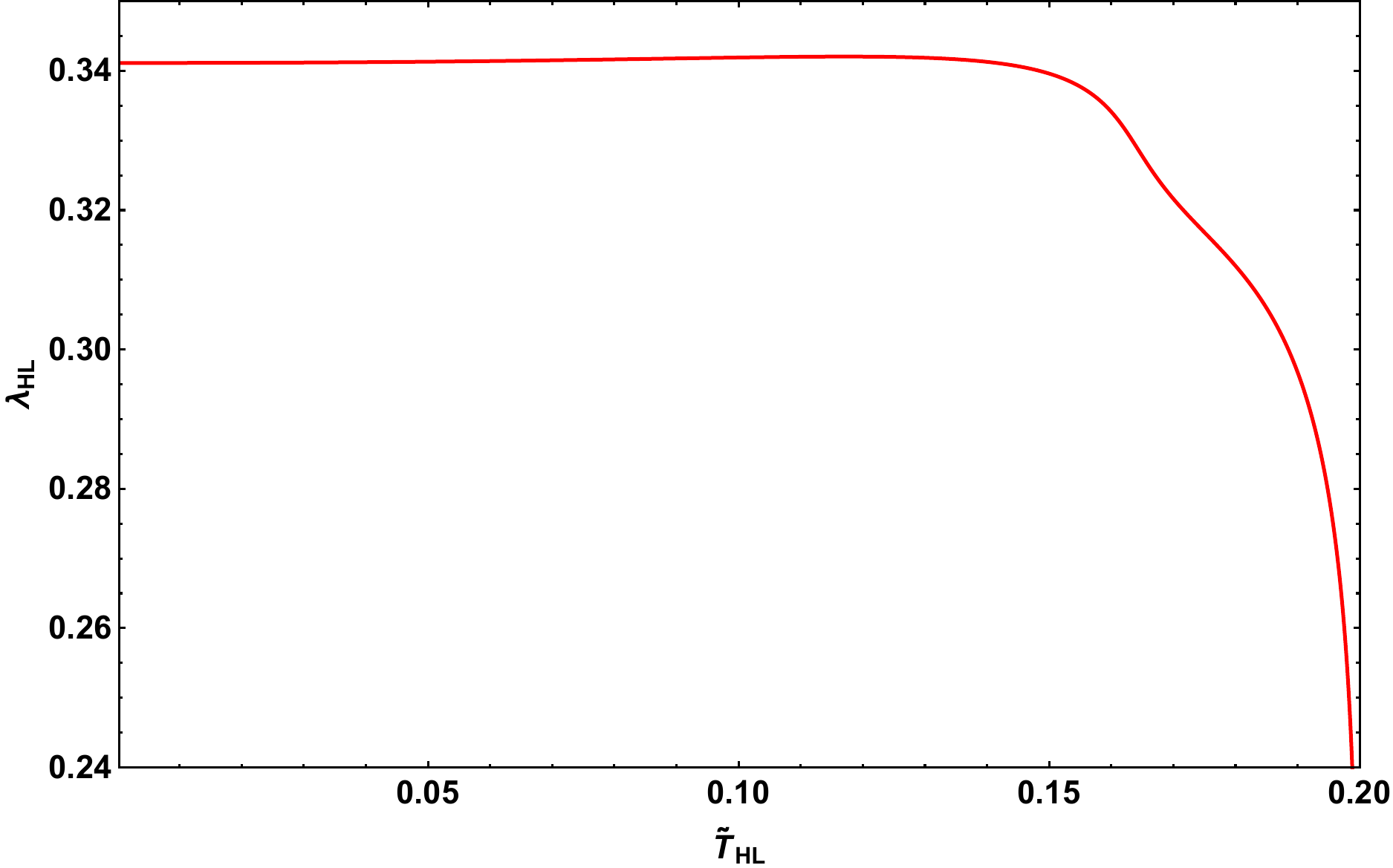}}
            \subfigure[$|\mathcal{G}_{HL}|-\tilde{T}_{HL}$]{
				\label{HLFT}
				\includegraphics*[scale=0.23]{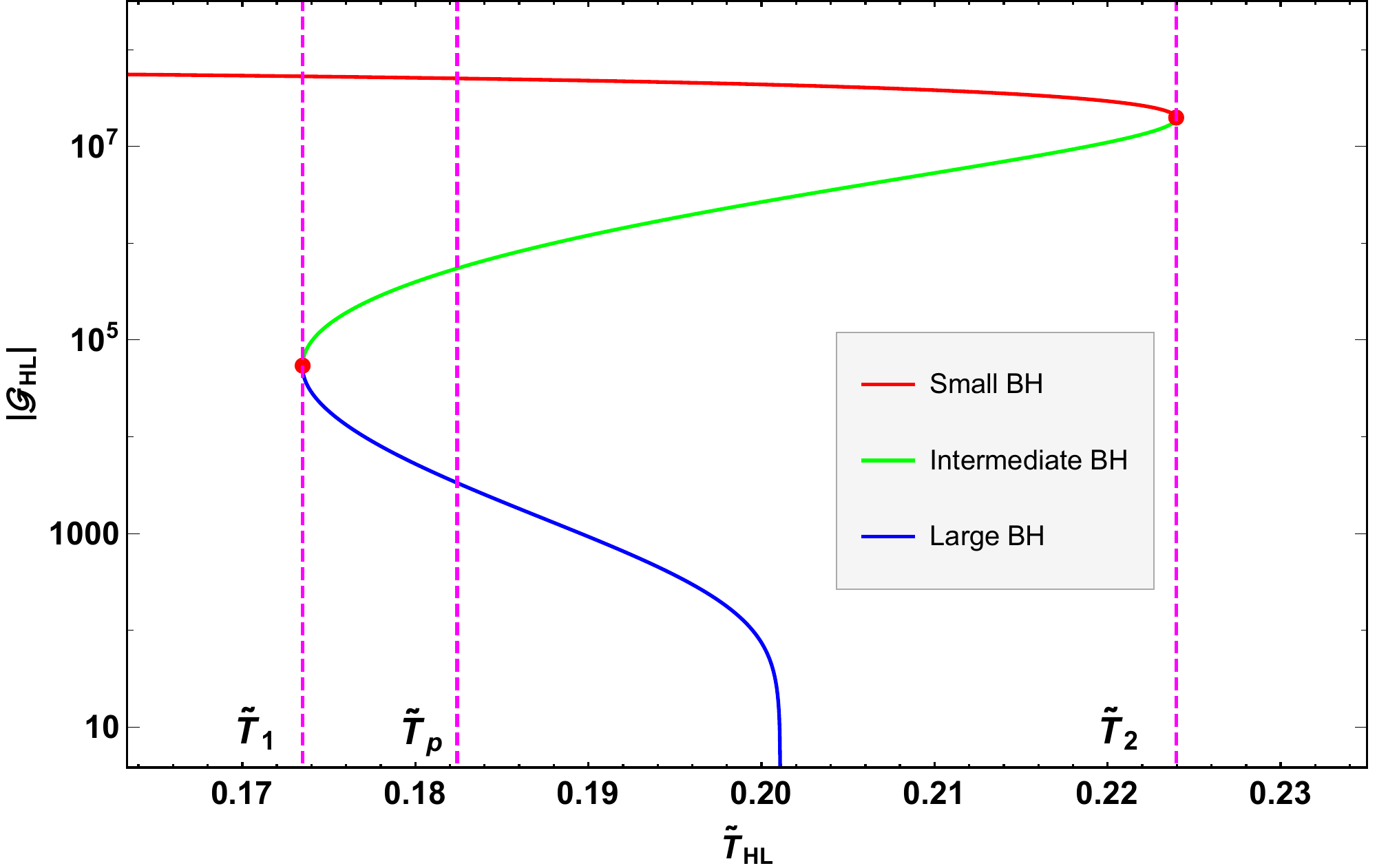}}
            \subfigure[$|\mathcal{G}_{HL}|-\tilde{T}_{HL}$]{
				\label{HLFT}
				\includegraphics*[scale=0.252]{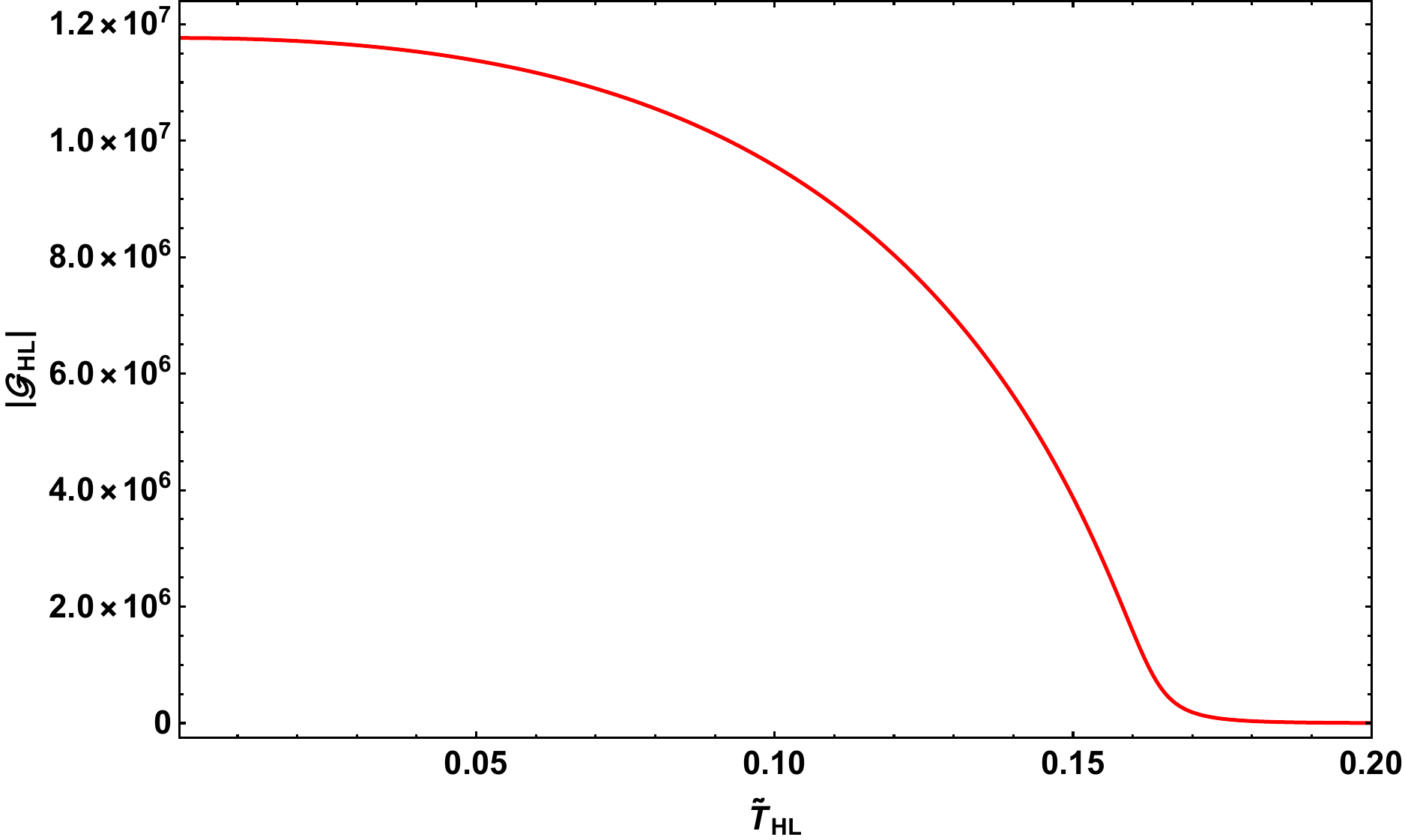}}
		\end{center}
		\caption{\label{fig1} Phase transition signatures in dynamics and geometry for timelike orbits, with $\tilde{g}_c=0.1042$, $a=0.6$, $L=20\ell$. Left column ($\tilde{g}=0.0615$, $\tilde{g}<\tilde{g}_c$): (a) free energy $\tilde{F}_{HL}$ versus $\tilde{T}_{HL}$, (c) Lyapunov exponent $\lambda_{HL}$ versus $\tilde{T}_{HL}$, (e) geometric quantity $|\mathcal{G}_{HL}|$ versus $\tilde{T}_{HL}$ (log scale). Right column ($\tilde{g}=0.1276$, $\tilde{g}>\tilde{g}_c$): (b) $\tilde{F}_{HL}$ versus $\tilde{T}_{HL}$, (d) $\lambda_{HL}$ versus $\tilde{T}_{HL}$, (f) $|\mathcal{G}_{HL}|$ versus $\tilde{T}_{HL}$ (log scale). The synchronized multivalued behavior of $\lambda_{HL}$ and $|\mathcal{G}_{HL}|$ in the spinodal region $\tilde{T}_{HL} \in (\tilde{T}_{1},\, \tilde{T}_{2})$ corresponds to the swallowtail structure in the free energy, with the phase transition occurring at $\tilde{T}_p$. All quantities become monotonic in the absence of a phase transition.
        }
	\end{minipage}
\end{figure*}

As shown in Fig.~\ref{fig1}, when a first-order phase transition occurs, as signified by the swallowtail structure in the $\tilde{F}_{HL}(\tilde{T}_{HL})$ curve ($\tilde{g}<\tilde{g}_c$), both the $\lambda_{HL}(\tilde{T}_{HL})$ and $|\mathcal{G}_{HL}(\tilde{T}_{HL})|$ curves exhibit multivalued behavior within the spinodal region $\tilde{T}_{HL} \in (\tilde{T}_{1}, \tilde{T}_{2})$. Crucially, in the absence of a phase transition ($\tilde{g}>\tilde{g}_c$), both $\lambda_{HL}(\tilde{T}_{HL})$ and $|\mathcal{G}_{HL}(\tilde{T}_{HL})|$ return to monotonic behavior, in precise agreement with the thermodynamic phase diagram. This demonstrates that both the geometric quantity $\mathcal{G}$ and the Lyapunov exponent $\lambda$ are linked to the first-order phase transitions of black holes. In particular, the geometric quantity $\mathcal{G}$, related to the extrinsic curvature of the MPS, serves as a robust probe for the first-order phase transition.

\section{Critical Exponents}
Critical exponents are used to describe the behavior of physical quantities of thermodynamic systems near the critical point and serve as important physical quantities for characterizing phase transitions. At the phase transition point, the small and large black hole branches coexist, leading to discontinuities in both the timelike Lyapunov exponent $\lambda$ and the geometric quantity $\mathcal{G}$. This section examines the continuity of $\lambda$ and $\mathcal{G}$ at the phase transition point $T_p$ for the Hayward–Letelier–AdS black hole, as well as their discontinuity near $T_p$. We investigate whether they can serve as order parameters and check whether their critical behavior aligns with the predictions of mean-field theory.

We define the timelike Lyapunov exponents for the small and large black hole branches within the spinodal region as $\lambda_S$ and $\lambda_L$, respectively, and the corresponding geometric quantities as $\mathcal{G}_S$ and $\mathcal{G}_L$. Their differences are defined as $\Delta \lambda$ and $\Delta \mathcal{G}$. When the critical point $T_p=T_c$, we have $\Delta \lambda=\Delta \mathcal{G}=0$. We will fit the numerical coefficients of $\Delta \lambda$ and $\Delta \mathcal{G}$ according to the following form \cite{PS01}
\begin{align}
\frac{\Delta\lambda}{\lambda_c}, \frac{\Delta \mathcal{G}}{\mathcal{G}_c} \sim \alpha(t-1)^\beta.
\end{align}
Here, $t=T_p/T_c$, $\alpha$ and $\beta$ are real numbers, the subscript ``$c$'' denotes the value of $\lambda$ and $\mathcal{G}$ at the critical point.

\begin{figure*}
	\begin{minipage}{1\hsize}
		\begin{center}
			
			\subfigure[]{
				\label{DL}
				\includegraphics*[scale=0.205]{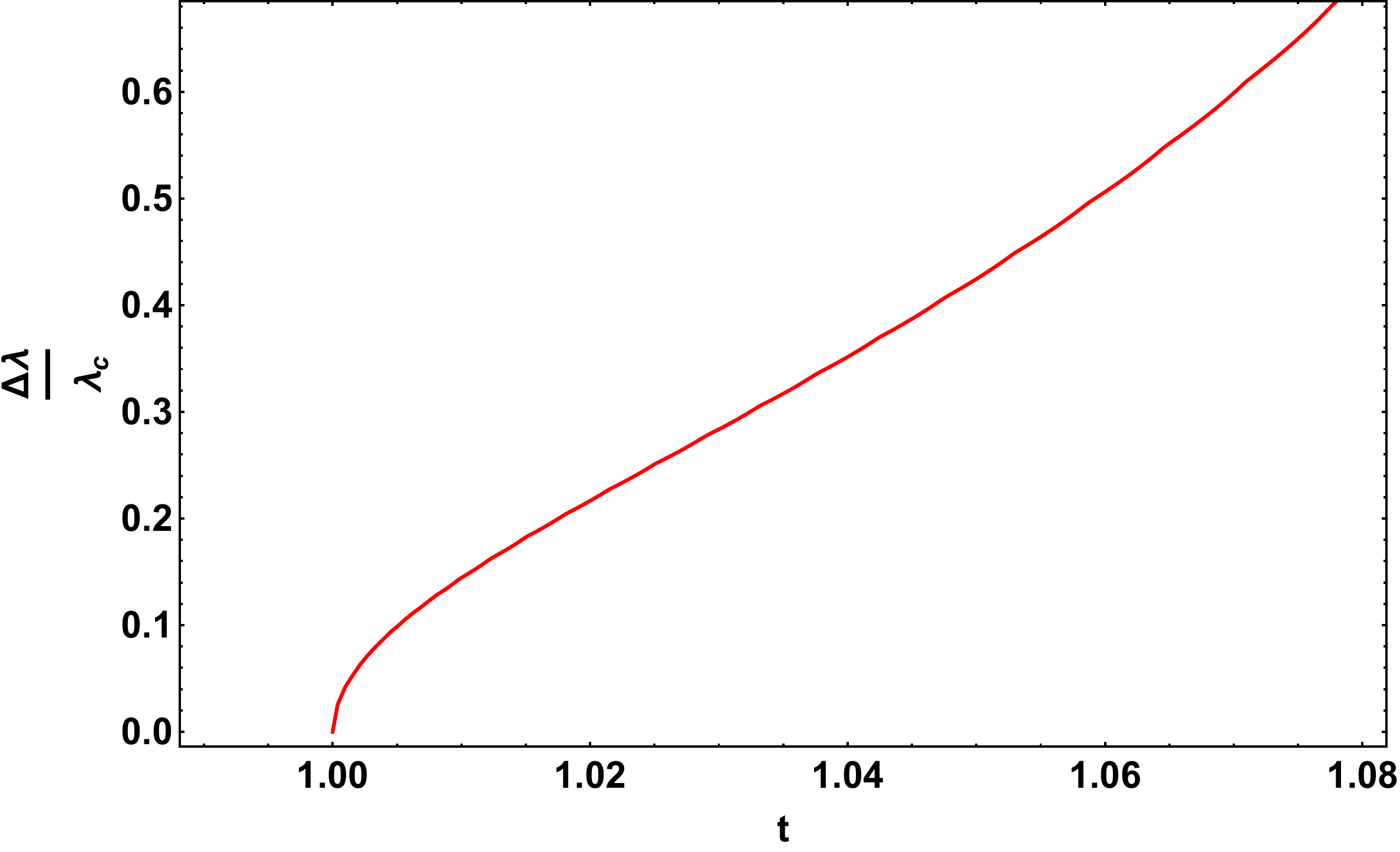}}
			\subfigure[]{
				\label{DK}
				\includegraphics*[scale=0.22]{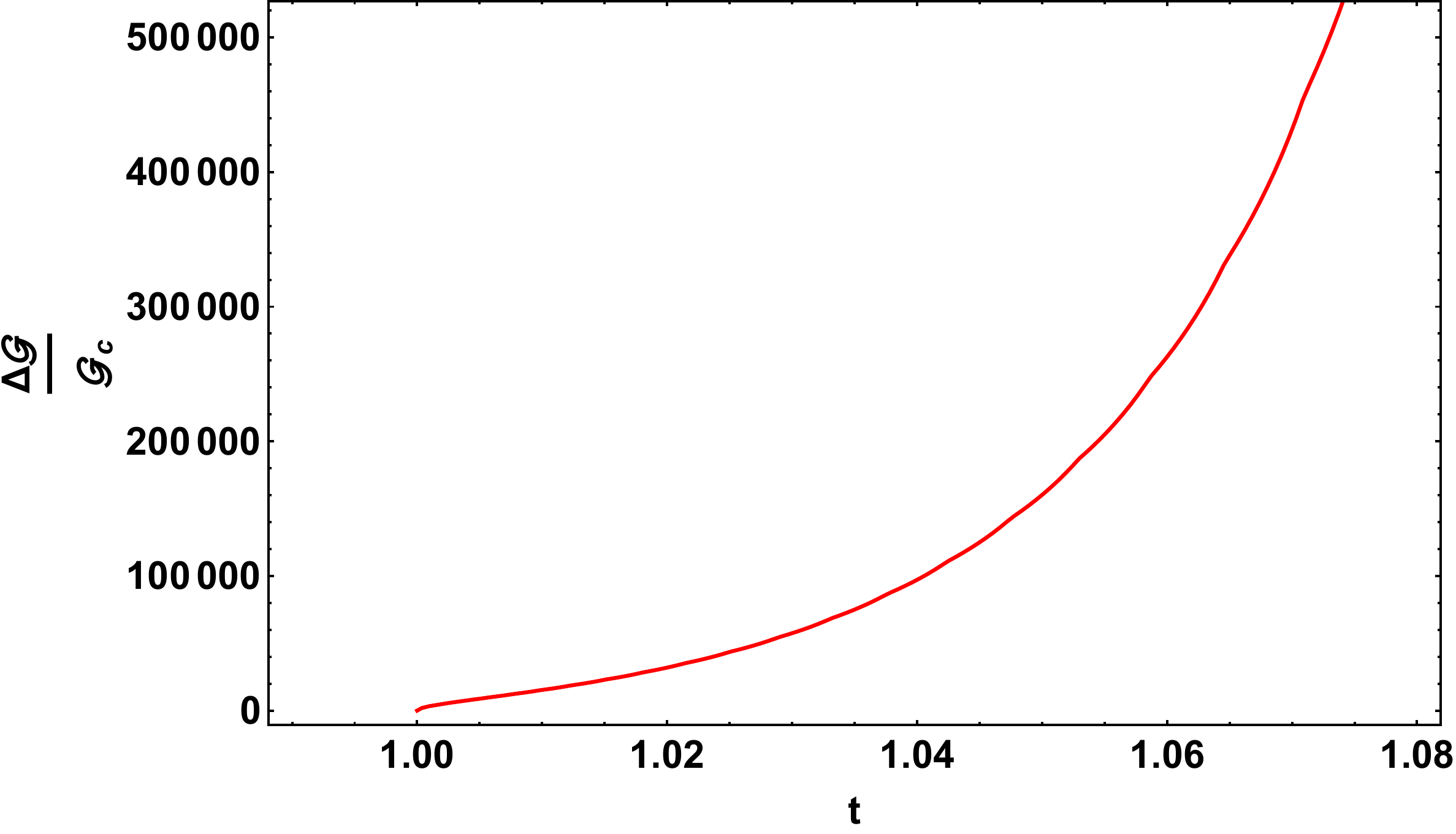}}
            
		\end{center}
		\caption{The reduced timelike Lyapunov exponents ${\Delta\lambda}/{\lambda_c}$ and geometric quantity ${\Delta \mathcal{G}}/{\mathcal{G}_c}$ for the coexistence small and large black holes. (a) ${\Delta\lambda}/{\lambda_c}$ and (b) ${\Delta \mathcal{G}}/{\mathcal{G}_c}$ versus reduced temperature $t$ for a Hayward-Letelier-AdS black hole.
        }
		\label{Fig.1}
	\end{minipage}
\end{figure*}

As shown in Fig.~\ref{Fig.1}, we plot the normalized differences ${\Delta\lambda}/{\lambda_c}$ and ${\Delta \mathcal{G}}/{\mathcal{G}_c}$ as functions of the reduced temperature $t$. When $T_p=T_c$, ${\Delta\lambda}/{\lambda_c}={\Delta \mathcal{G}}/{\mathcal{G}_c}=0$. When $T_p$ deviates from $T_c$, both ${\Delta\lambda}/{\lambda_c}$ and ${\Delta \mathcal{G}}/{\mathcal{G}_c}$ take non-zero values, indicating the emergence of discontinuity. Finally, by fitting the curves of $\Delta\lambda/\lambda_c(t)$ and $\Delta\mathcal{G}/\mathcal{G}_c(t)$, we obtain the following scaling behavior
\begin{align}
\frac{\Delta\lambda}{\lambda_c} \simeq 10^{0.8286}(t-1)^{0.5918},
\end{align}
and
\begin{align}
\frac{\Delta \mathcal{G}}{\mathcal{G}_c} \simeq 10^{14.6054}(t-1)^{1.0244}.
\end{align}
These results indicate that the critical exponents of ${\Delta\lambda}/{\lambda_c}$ ($0.5918$) and ${\Delta \mathcal{G}}/{\mathcal{G}_c}$ ($1.0244$) deviate from the value of $1/2$ observed in the Reissner-Nordström black hole, suggesting that regular black holes exhibit richer critical behavior in their geometry and dynamics than their singular counterparts. Given Eq.~(\ref{G}), the geometric quantity $\mathcal{G}$ is proportional to $\lambda^2$, but the proportionality coefficient involves metric functions. Therefore, the critical exponent of ${\Delta \mathcal{G}}/{\mathcal{G}_c}$ is not simply twice that of ${\Delta\lambda}/{\lambda_c}$ ($1.1836$), but is instead influenced by the scaling of the coefficient itself. This leads to a deviation of the critical exponent of ${\Delta \mathcal{G}}/{\mathcal{G}_c}$ from the expected value $1.1836$. This result further reveals the complex coupling between geometric and dynamic quantities in regular black holes.

\section{Conclusion}
In this work, we have systematically extended the correspondence between geometry and dynamics from unstable null orbits to the timelike case and uncovered their fundamental distinction. Within the framework of the MPS theory, we constructed a new geometric quantity $\mathcal{G}$ and demonstrated that $\mathcal{G} \propto -\lambda^2$ holds for unstable timelike orbits, where $\lambda$ is the Lyapunov exponent characterizing the orbital chaos. This finding fills a crucial theoretical gap by establishing this correspondence specifically for massive particles.

Moreover, we have found that the geometric quantity $\mathcal{G}$ exhibits multivalued behavior near the black hole first-order phase transition, synchronizing with the Lyapunov exponent $\lambda$. This reveals that spacetime geometry itself encodes thermodynamic phase transition information, thereby establishing a geometric framework for probing black hole thermodynamics in the timelike case. Furthermore, we find that the critical behaviors of both $\mathcal{G}$ and $\lambda$ deviate from the mean-field expectation, indicating that the geometric and dynamic properties of regular black holes differ from those of their singular counterparts. This result also provides a basis for further investigation into this distinction.

Our work demonstrates that thermodynamic phase transitions are detectable via a geometrical spacetime quantity. This provides new insights into the fundamental question of how spacetime geometry encodes black hole thermodynamics and lays the groundwork for exploring the deep geometry-thermodynamics connection in other gravitational theories.

\section*{Acknowledgments}
This work was supported by the National Natural Science Foundation of China (Grants Nos. 12533001, 12575049, and 12473001), the National SKA Program of China (Grants Nos. 2022SKA0110200 and 2022SKA0110203), the China Manned Space Program (Grant No. CMS-CSST-2025-A02), and the 111 Project (Grant No. B16009).

\bibliography{reference}

\end{document}